\documentclass[a4paper,11pt,dvips]{article}
\usepackage{graphicx}
\usepackage{amsmath}
\usepackage{amssymb}
\usepackage{latexsym}
\usepackage{color}

\newcommand{\bra}{\begin{array}}
\newcommand{\era}{\end{array}}
\newcommand{\beq}{\begin{equation}}
\newcommand{\eeq}{\end{equation}}
\newcommand{\beqar}{\begin{eqnarray}}
\newcommand{\eeqar}{\end{eqnarray}}

\def\BC{\bb C}
\def\_\BC{\bbi C}

\def\( {{\left(}}
\def\) {\right)}
\def\[ {{\left[}}
\def\] {{\right]}}
\def\no2 {{\textstyle{n\over 2}}}

\newcommand{\om}{\omega}

\newcommand{\si}{\sigma}

\newcommand{\va}{\varepsilon}

\newcommand{\pa}{\partial}

\newcommand{\del}{\delta}

\newcommand{\da}{\dagger}

\newcommand{\lb}{\label}

\newcommand{\PR}[1]{ {\it Phys.~Rev.} {\bf #1}}
\newcommand{\PRL}[1]{ {\it Phys.~Rev.~Lett.} {\bf #1}}

\begin{document}
\begin{titlepage}
\setcounter{page}{1}
\renewcommand{\thefootnote}{\fnsymbol{footnote}}

\begin{flushright}
ucd-tpg:1103.04\\
\end{flushright}

\vspace{5mm}
\begin{center}

{\Large \bf {Massless Dirac Fermions in Electromagnetic  Field}}

\vspace{5mm}

{\bf Ahmed Jellal$^{a,b,c}$\footnote{\sf ajellal@ictp.it - jellal.a@ucd.ca.ma}},
{\bf Abderrahim El Mouhafid$^{b,c}$} 
and {\bf Mohammed Daoud$^{d,e}$} 

\vspace{5mm}

{$^a$\em Physics Department, College of Sciences, King Faisal University,\\
PO Box 380, Alahsa 31982, Saudi Arabia}

{$^b$\em Saudi Center for Theoretical Physics, Dhahran, Saudi Arabia}

{$^{c}$\em Theoretical Physics Group,  
Faculty of Sciences, Choua\"ib Doukkali University},\\
{\em PO Box 20, 24000 El Jadida,
Morocco}

$^{d}${\em Max Planck Institute for the Physics of Complex Systems,
\\ 01187 Dresden, Germany}

{$^{e}$\em Department of Physics,
Faculty of Sciences, Ibn Zohr University},\\
{\em PO Box 8106,  80006 Agadir,
Morocco}


\vspace{3cm}

\begin{abstract}

We study the relations between massless Dirac fermions  in an electromagnetic field
and atoms in quantum optics.
After getting the solutions of the energy spectrum,
we show that it is possible to reproduce the 2D Dirac Hamiltonian, with all its quantum
relativistic effects, in a controllable system as a single trapped ion
through the Jaynes--Cummings and anti-Jaynes--Cummings models.
Also we show that under certain conditions
the evolution of the Dirac Hamiltonian provides us with %
Rashba spin-orbit and
linear Dresselhaus couplings.
Considering the multimode multiphoton Jaynes-Cummings model
interacting with $N$ modes of
electromagnetic field prepared in general pure quantum states,
we analyze the Rabi oscillation.
Evaluating time evolution of the Dirac position operator,
we determine the Zitterbewegung frequency and the corresponding
oscillating term {as function of the electromagnetic field}.

\end{abstract}
\end{center}
\end{titlepage}

\newpage

\section{Introduction}

%

In graphene~\cite{1111,2222, goerbig1}, the linear electronic band
dispersion near the Dirac points gave rise to charge carriers (electrons or holes) that
propagate as if they were massless fermions with speeds of the order of $10^6 m/s$ rather
than the speed of light $3\times 10^8 m/s$. Hence, charge carriers in this structure should be
described by the massless Dirac equation rather than the usual Schr\"odinger equation.
The physics of relativistic electrons is thus now experimentally accessible in graphene
based
solid-state devices, whose behavior differs drastically from that of similar devices
fabricated with usual semiconductors. Consequently, new unexpected phenomena have
been observed while other phenomena that were well-understood in common semiconductors,
such as the quantum Hall effect~\cite{1,2}.
Thus, graphene devices enabled the study of relativistic dynamics
in controllable nano-electronic circuits (relativistic electrons on-a-chip) and their
behavior probes our most basic understanding of electronic processes in solids. It also
allowed to establish a relationship between massless Dirac fermions in two dimensions
and quantum optics, for instance one can see~\cite{ziegler,mishchenko}.

On the other hand,
Jaynes--Cummings (JC) model is describing
the basic interaction of a two-level atom and
quantized  field, which  is also the cornerstone for the
treatment of the interaction between light and matter in quantum
optics~\cite{4}. It can be used to explain many quantum phenomena,
such as the collapses and revivals of the atomic population
inversions, squeezing of the quantized field and the atom-cavity
entanglement. Recent experiments showed that JC model
can be implicated in quantum-state engineering and quantum
information processing, e.g. generation of Fock states~\cite{5}
and entangled states~\cite{6}, and the implementations of quantum
logic gates~\cite{7}, etc. Originally, JC model is physically
implemented with a cavity quantum electrodynamics system,
see for instance~\cite{8}.
Certainly, there has been also interest to
realize JC model with other physical
systems. A typical system is a cold ion trapped in a Paul trap and
driven by classical laser beams~\cite{9,101} where the
interaction between two selected internal electronic levels and
the external vibrational mode of the ion can be induced.

The relationship between the relativistic systems and
atoms in quantum optics is studied for differen purposes.
{With this respect, the first theoretical proposal was done by Lamata {\it et al.} \cite{lamata1}
and its experimental realization was done by Gerritsma {\it et al.} \cite{gerritsma1}. Subsequently, different
progresses appreared on the subject \cite{casanova,gerritsma2,lamata2,11} where for}
 example in~\cite{11},
the dynamics of the 2+1 Dirac oscillator exactly was studied and spin oscillations due to a Zitterbewegung
of purely relativistic origin was found. An exact mapping of this quantum-relativistic system onto a JC model 
is established. This
equivalence allowed  to map a series of quantum optical phenomena onto the relativistic oscillator and vice
versa. A realistic experimental proposal was made, in reach with current technology, for studying the equivalence
of both models using a single trapped ion.

Motivated by different developments  and in particular~{\cite{lamata1,11}},
we consider massless Dirac fermions in
an electromagnetic field and  show
its link with
JC and anti-Jaynes--Cummings (AJC) models. 
This can be done by determining the solutions of the energy spectrum and mapping
the corresponding eigenspinors in terms of two states of JC model.
Furthermore,
we analyze the Rabi oscillations
by considering atoms {coupled to the field states and  evolve} in time
according to the stationary eigenspinors. This allows us to
obtain two normalization coefficients in terms of
the Rabi frequency, which verify the normalization condition.
By evaluating the probabilities of the atom in
the ground and excited
states, we show that
the probability predicts
sinusoidal Rabi oscillations, much like the classical case,
and exists
even when the field is initially in the ground state.

{Finally}, we deal with another aspect of Dirac fermions in graphene~\cite{katsnelson,rusin,goerbig}
that is the Zitterbewegung (ZB) effect. Indeed, by calculating time evolution of the Dirac position
operator, we
end up with an extra term. After inspection, we show that
this term  represents a quantum oscillation about the classical motion and
indicates the existence of  the ZB effect. By evaluating the expectation value of the Dirac position,
we obtain the ZB frequency in terms of the energy levels and magnetic field, which disappears 
{for a null energy.} 

The present paper is organized as follows. In section 2, we
consider the Hamiltonian describing one massless Dirac fermion in electromagnetic field
with Zeeman effect. By choosing the Landau gauge, we obtain the eigenvalues and
the corresponding eigenspinors.
After setting different ingredients, we establish the link between our system and quantum optics
through JC and AJC models in section 3.
The Rabi oscillations for atoms coupled to field states in graphene will be analyzed
in section 4. Using the Heisenberg picture dynamics we study
the ZB effect and determine its oscillation frequency  in section 5. 
{Finally we close our work
by concluding and giving some perspectives.} 


\section{Dirac fermions in electromagnetic field}

In graphene, the
two Fermi points, each with a two-fold band degeneracy, can be described by a low-energy continuum
approximation with a four-component envelope wavefunction whose components are labeled by a
Fermi-point pseudospin $=\pm 1$ and a sublattice forming an honeycomb. 
Under this approximation we can describe our charge carriers, in electromagnetic field $(\vec E, \vec
B)$ where  $\vec E= E \vec e_y$,  by a gauge
invariant Dirac Hamiltonian in 2+1 space-time dimensions with minimal coupling to the vector
potential and Zeeman effect
as follows
\begin{equation}\lb{1}
H_{D}=v_F \vec{\sigma}\cdot \vec{\pi}+\sigma_y
eEy+\frac{1}{2}g\mu_BB\sigma_{z}
\end{equation}
where {$v_F\approx 10^{6}m/s$ is the Fermi velocity, the conjugate momentum is
$\vec\pi=\vec p-\frac{e}{c}\vec A$}, $g$ is the Land\'e factor, $\mu_B$ is the Bohr magneton,
{$\vec\si$ are Pauli spin matrices
 and
the nonzero matrix elements of $\sigma_{z}$}
are $\langle\uparrow\mid\sigma_{z}\mid\uparrow\rangle=+1$ and
$\langle\downarrow\mid\sigma_{z}\mid\downarrow\rangle=-1$.
Choosing the vector potential in the Landau gauge
$\vec{A}(x,y)=(-By,0)$ 
to write (\ref{1}) as
\begin{eqnarray}
H_{D} =  v_F\left(%
\begin{array}{cc}
  0 & p_x-ip_y-{eB\over c}y -i\frac{eE}{v_F}y \\
  p_x+ip_y-{eB\over c}y+ i\frac{eE}{v_F}y   & 0 \\
\end{array}%
\right) + \frac{1}{2}g\mu_BB\sigma_{z}. 
\end{eqnarray}
By setting $\xi=1 +i{cE\over {v_F} B}$ and $\bar\xi=1-i{cE\over {v_F} B}$, we
obtain the form
 \begin{equation}\label{al}
H_D=v_F\left(%
\begin{array}{cc}
  0 & p_x-ip_y-{\hbar\over l_{B}^{2}}\ \xi y\\
  p_x+ip_y- {\hbar\over l_{B}^{2}}\ \bar\xi y & 0 \\
\end{array}%
\right)+ \frac{1}{2}g\mu_BB\sigma_{z}
\end{equation}
where  
 $l_B=\sqrt\frac{\hbar c} {eB }$ is the magnetic length.

To diagonalize the above Hamiltonian, it is convenient to introduce the
annihilation and creation operators. They are
\begin{equation}\lb{real}
a= \frac{l_B}{\hbar\sqrt{2}}\left(ip_x -p_y -i\frac{\hbar}{l_B^2} \xi y\right) ,\qquad
a^{\dagger}=\frac{l_B}{\hbar\sqrt{2}}\left( -ip_x -p_y +i\frac{\hbar}{l_B^2} \bar\xi y \right)
\end{equation}
which satisfy the commutation relation
\beq
\left[a^{\dagger},a \right]=\mathbb{I}.
\eeq
These can be used to write $H_D$ as
\begin{equation}\label{99}
H_{D}=\left(%
\begin{array}{cc}
  \frac{1}{2}g\mu_BB & i\hbar\omega_{c}a^{\da} \\
  -i\hbar\omega_{c}a & -\frac{1}{2}g\mu_BB  \\
\end{array}%
\right)
\end{equation}
where we have set the frequency $\omega_{c}=\sqrt{2}\frac{v_{F}}{l_{B}}$.

The complete solutions of the energy spectrum  can be obtained by solving the time independent
Dirac equation  $H_{D}\phi=\va\phi$.
{Note that, the form of our Hamiltonian suggests to
separate variables by writing the eigenspinors as}
\beq
{\phi(x,y)=\chi(x)\psi(y), \qquad \psi(y)= \left(\psi_{1},\psi_{2}\right)^{t}, \qquad
 \chi (x) = e^{ik_x x}}
\eeq
with $k_x$ is a real parameter that stands for the wave number of the excitations along
$x$-direction. Thus, the reduced eigenvalue equation is {given by}
\begin{eqnarray}\label{9}
H_{D}\left(\begin{array}{c}
  \mid \psi_{1}\rangle \\
  \mid \psi_{2}\rangle \\
\end{array}\right)=\va\left(\begin{array}{c}
  \mid \psi_{1} \rangle \\
  \mid \psi_{2} \rangle \\
\end{array}\right)
\end{eqnarray}
which {leads to the relations}
\begin{eqnarray}
i\hbar\omega_{c}a^{\da}\mid \psi_{2} \rangle =\left(\va-\frac{1}{2}g\mu_{B}B\right)\mid \psi_{1} \rangle \label{10}\\
-i\hbar\omega_{c}a \mid \psi_{1} \rangle =\left(\va+\frac{1}{2}g\mu_{B}B\right) \mid \psi_{2} \rangle. \label{11}
\end{eqnarray}
{As we will see soon these will be solved by making separation between
the positive and negative energy solutions
as well as discussing the relation between them.}

The positive energy solution can be obtained
from {(\ref{10}) and (\ref{11})} 
to get the following relation between spinor components
\begin{eqnarray}\label{131}
\mid \psi_{2} \rangle =\frac{-i\hbar\omega_{c}}{\left(\va+\frac{1}{2}g\mu_{B}B\right)}a \mid \psi_{1} \rangle.
\end{eqnarray}
Clearly, 
it is necessary to impose the condition
$\va\neq-\frac{1}{2}g\mu_{B}B<0$, which tells us that (\ref{131}) is valid only for
positive energies corresponding to the positive eigenspinors denoted by
$\mid \psi^{+} \rangle=\left(%
\begin{array}{c}
  \mid \psi_{1}^{+} \rangle \\
 \mid \psi_{2}^{+} \rangle  \\
\end{array}%
\right)$.
{Now, from (\ref{131}) we} show that  $\mid \psi_{1}^{+} \rangle$ 
satisfies the second differential
equation 
\begin{eqnarray}\label{13}
\left[\va^{2}-\left(\frac{1}{2}g\mu_{B}B\right)^{2}-\hbar^{2}\omega_{c}^{2}a^{\da}a\right] \mid \psi_{1}^{+} \rangle=0
\end{eqnarray}
which is nothing but a simple harmonic oscillator type. Thus, 
the positive eigenvalues  are given
by
\begin{eqnarray}\label{13}
\lb{engy}
\va_{n}^{+}=+\sqrt{\hbar^{2}\omega_{c}^{2}n+\frac{1}{4}(g\mu_{B}B)^{2}}, \qquad n={0, 1, 2, \cdots}
\end{eqnarray}
where the corresponding eigenstates 
take the form
\begin{eqnarray}\label{14}
\mid \psi_{1}^{+} \rangle = \mid n\rangle = {\frac{\left(a^{\da}\right)^n}{\sqrt{n!}} \mid 0\rangle}.
\end{eqnarray}
{At this level, it is interesting to mention that these eigenstates are clearly depending on
the electromagnetic
field. To clarify this statement, we write them as 
\begin{eqnarray}\label{144}
\mid \psi_{1}^{+} \rangle =  \frac{\left({-il_B}\right)^n}{\sqrt {2^n n!}}
 \left(  k_x +\pa_y -\frac{\hbar}{l_B^2} \bar\xi y \right)^{n} \mid 0\rangle
\end{eqnarray}
with $k_x$ is the wave vector along $x$-direction and the parameter $\bar\xi= 1- i\frac{cE}{v_F B}$
is a function of the ratio between the two fields.}
Now inserting (\ref{14}) into  (\ref{131}) to find
\begin{eqnarray}
\mid \psi_{2}^{+} \rangle
=\frac{-i\hbar\omega_{c}}{\va_{n}^{+}+\frac{1}{2}g\mu_{B}B}a\mid
n\rangle=-
i\sqrt{\frac{\va_{n}^{+}-\frac{1}{2}g\mu_{B}B}{\va_{n}^{+}+\frac{1}{2}g\mu_{B}B}}\mid
n-1\rangle.
\end{eqnarray}
Combining all, we write
the eigenspinors as
\begin{eqnarray}\label{16}
\mid \psi_{n}^{+} \rangle=\left(%
\begin{array}{c}
  u_{n}^{+}\mid n\rangle \\
-i u_{n}^{-}\mid n-1\rangle \\
\end{array}%
\right)\
\end{eqnarray}
where the amplitude  $u_{n}^{\pm}$ read as
\begin{eqnarray}\label{upm}
u_{n}^{\pm}=\left(\frac{1}{2}\pm\frac{\frac{1}{4}g\mu_{B}B}{\sqrt{\hbar^{2}\omega_{c}^{2}n+\frac{1}{4}(g\mu_{B}B)^{2}}}\right)^{\frac{1}{2}}.
\end{eqnarray}

As far as the negative energy solution is concerned, we use similar approach to end up with
the eigenvalues
\begin{eqnarray}
\lb{engy}
\va_{m}^{-}=-\sqrt{\hbar^{2}\omega_{c}^{2}(m+1)+\frac{1}{4}(g\mu_{B}B)^{2}}, \qquad m={0, 1, 2, \cdots}.
\end{eqnarray}
To derive the corresponding negative eignespinors, we use
the energy conservation
\begin{eqnarray}
(\va_{n}^{+})^{2}=(\va_{m}^{-})^{2}
\end{eqnarray}
which tells us that 
the two quantum numbers
are related by $n=m+1$. Thus, we obtain
\begin{eqnarray}\label{19}
\mid \psi_{n}^{-} \rangle =\left(%
\begin{array}{c}
  +iu_{n}^{+}\mid n+1\rangle \\
  u_{n}^{-}\mid n\rangle \\
\end{array}%
\right)
\end{eqnarray}
where $u_{n}^{\pm}$ are given in (\ref{upm}). {We recall that all eigenspinors obtained so far are electromagnetic field $(\vec E, \vec B)$ dependent.}

For later use, it is convenient to write the eigenspinors (\ref{16}) and (\ref{19}) in an appropriate form.
This can be done by using the Pauli spinors, such as  
\begin{eqnarray}
\mid \psi_n^{+}\rangle\equiv\mid \psi_{n,\si,\va_{n}^{+}}\rangle  &=& u_{n}^{+}\mid n,\uparrow\rangle -iu_{n}^{-}\mid n-1,\downarrow\rangle \label{aaa}\\
\mid \psi_n^{-}\rangle\equiv \mid \psi_{n,\si, \va_{n}^{-}} \rangle &=& u_{n}^{-}\mid n,\uparrow\rangle
+iu_{n}^{+}\mid n-1,\downarrow\rangle \label{bbb}.
\end{eqnarray}
This summarizes the obtained results so far, which  will
be employed to study different issues in the forthcoming anaylsis. More precisely,
we will implement them to establish a link between
massless Dirac Fermions
in  electromagnetic field and
atoms
in quantum optics. Subsequentely, we will discuss the possibility to
make contact with the Rabi oscillations and ZB effect as well.

\section{Link with Jaynes--Cummings models} 

As claimed before, we would like to
use our results in order to make contact with quantum optic systems
through two relevant models, i.e.  JC and AJC. In doing so, first let us  prepare our system to make the required
contact easy and clear. From (\ref{10}) and (\ref{11}) we show that
the Hamiltonian (\ref{99}) takes the form
\beq
H_{D}=i\hbar
\omega_{c}\left(a^{\da}\mid\psi_{2}\rangle\langle\psi_{1}\mid-
a\mid\psi_{1}\rangle\langle\psi_{2}\mid\right)+\frac{1}{2}g\mu_BB\sigma_{z}
\eeq
which can easily be written as
\begin{equation}\label{33}
H_{D} = \hbar \left(g'\sigma
^{-}a^{\da}+g'^{*}\sigma^{+}a\right)+\frac{1}{2}g\mu_BB\sigma_{z}.
\end{equation}
This Hamiltonian can be interpreted as a model of quantum optics
where $g'=i\omega_{c}$ is the coupling strength
between the atom and electromagnetic field,
$\sigma^{+}$ and $\sigma^{-}$ are the spin raising and
lowering operators, respectively. More precisely, 
(\ref{33}) 
is describing a
single two-state atom, represented by the Pauli matrices, 
interacting with a (single-mode quantized) electromagnetic
field. 
This statement will be clarified in the next.

We will show how to implement the dynamics of (\ref{33}) in a
single ion inside a Paul trap with two frequencies $\nu_{x}$ and
$\nu_{y}$. 
For this, let us  consider a system of  two-level atom of
  the ground  $\mid e\rangle$ and excited  $\mid
f\rangle$ states where
the eigenspinors (\ref{16}) can be split into
\begin{eqnarray}\label{27}
\mid\psi\rangle=\mid\psi_{1}\rangle\mid
e\rangle+\mid\psi_{2}\rangle\mid f\rangle.
\end{eqnarray}
These two 
metastable internal states $\mid e\rangle$
and $\mid f\rangle$ are  corresponding to the following energies
\beq
\va_{n_{\sf e}}=\hbar\omega_{n_{\sf e}}= \hbar \sqrt{\omega_{c}^{2}n_{e}+(1/2g\mu_{B}B/\hbar)^{2}}, \qquad
\va_{n_{\sf f}}=\hbar\omega_{n_{\sf f}}= \sqrt{\omega_{c}^{2}n_{f}+(1/2g\mu_{B}B/\hbar)^{2}}
\eeq
where the energy difference between them is characterized by the
transition frequency $\omega_{0}= 
\left(\va_{n_{\sf e}}-\va_{n_{\sf f}}\right)/\hbar$. 
All observable quantities of this two-state system can be conveniently
represented by the Pauli operators, such as 
\begin{eqnarray}
\sigma_{z}=\mid e\rangle\langle e\mid-\mid f\rangle\langle f\mid,
& \qquad  \sigma^{+}=\mid e\rangle\langle f\mid,  & \qquad
\sigma^{-}=\mid f\rangle\langle e\mid
\end{eqnarray}
where the expectation value of the operator $\sigma_{z}$ is the atomic
inversion, while $\sigma^{+}$ and $\sigma^{-}$ induce upward and
downward transition respectively. It is sometimes convenient to
work with the Hermitian combinations of $\sigma^{+}$ and $\sigma^{-}$,
i.e.
$\sigma_{x}=\sigma^{+}+\sigma^{-}$ and 
$\sigma_{y}=i(\sigma^{+}-\sigma^{-})$.
Using these
 to write the Hamiltonian (\ref{33}) as
\begin{eqnarray}\label{3h}
H_{D}=\upsilon_{F} \left(\sigma_{x}p_{x}+\sigma_{y}p_{y}\right)
+\frac{\hbar\omega_{c}}{\sqrt{2}l_{B}}\left(-\sigma_{x}+  
\frac{c E}{B}
\sigma_{y}\right)y+\frac{1}{2}g\mu_BB\sigma_{z}
\end{eqnarray}
which is showing its dependence on the electric field through the extra term $\frac{c E}{B}
\sigma_{y}$ appearing in the Hamiltonian form. Clearly
for $E=0$, one recovers a system of massless Dirac fermions in magnetic field 
and thus the extra  term can be seen as a shift to the spin $\si_x$. Certainly,
this will play a crucial role in the forthcoming analysis and make difference with respect
to the former results obtained in this direction~\cite{11}.

Having set all ingredients, let us  show that
the Dirac  Hamiltonian (\ref{3h}) can be linked with JC and AJC models. We start by introducing
JC model, usually called red-sideband excitation, which is consisting of a laser field acting resonantly on two
internal levels. Typically, a resonant JC coupling induces an
excitation in the internal levels while producing a deexcitation
of the motional harmonic oscillator and viceversa. The resonant
JC Hamiltonian is given by
\begin{equation}\label{3gh}
    H_{i}^{JC}=\hbar\eta_{i}\Omega_{i} \left(\sigma
^{+}a_{i}e^{i\phi}+\sigma^{-}a^{+}_{i}e^{-i\phi}\right)+\hbar\delta_{i}\sigma_{z}
\end{equation}
where $a_{i}$ and $a_{i}^{+}$ $(i=x, y)$ are the annihilation
and creation operators associated with a motional degree of
freedom. $\eta_{i}:=k_{i}\sqrt{\hbar/2M\nu_{i}}$ is the Lamb-Dicke
parameter~\cite{101}, $k_{i}$ is the wave number of the
driving field, $M$ is the ion mass, $\nu_i$ are the natural trap
frequencies,  $\phi$ is the red-sideband phase,  $\Omega_{i}$ and $\delta_{i}$ are the excitation coupling strengths.
Using the realization (\ref{real}) for the annihilation and creation operators to map (\ref{3gh}) as
\begin{eqnarray}\label{q}
H_{i}^{JC} &=&\frac{\eta_{i}\Omega_{i}l_{B}}{\sqrt{2}}\left[i(\sigma^{+}e^{i\phi}-
\sigma^{-}e^{-i\phi})p_{x}-(\sigma^{+}e^{i\phi}+\sigma^{-}e^{-i\phi})p_{y}\right]\nonumber \\
&& +\frac{\hbar\eta_{i}\Omega_{i}}{\sqrt{2}l_{B}}\left[-i(\sigma^{+}e^{i\phi}-\sigma^{-}e^{-i\phi})-
\frac{c E}{B}(\sigma^{+}e^{i\phi}+\sigma^{-}e^{-i\phi})\right]y+\hbar\delta_{i}\sigma_{z}.
\end{eqnarray}
On the other hand, the AJC Hamiltonian reads as
\beq
\label{32}
H_{i}^{AJC} = \hbar\eta_{i}\Omega_{i} \left(\sigma
^{+}a_{i}^{+}e^{i\varphi}+\sigma^{-}a_{i}e^{-i\varphi}\right)
\eeq
where $\varphi$ is the blue-sideband phase. In similar way to (\ref{3gh}), we show that
$H_{i}^{AJC}$ takes the form
\begin{eqnarray} \label{32}
 H_{i}^{AJC}  &=&\frac{\eta_{i}\Omega_{i}l_{B}}{\sqrt{2}}\left[-i(\sigma^{+}e^{i\varphi}-\sigma^{-}e^{-i\varphi})p_{x}-
(\sigma^{+}e^{i\varphi}+\sigma^{-}e^{-i\varphi})p_{y}\right]\\
&& +\frac{\hbar\eta_{i}\Omega_{i}}{\sqrt{2}l_{B}}\left[i(\sigma^{+}e^{i\varphi}-\sigma^{-}e^{-i\varphi})-
\frac{c E}{B}(\sigma^{+}e^{i\varphi}+\sigma^{-}e^{-i\varphi})\right]y\nonumber.
\end{eqnarray}

By fixing different parameters,
let us see how does look like the sum of $H^{JC}_{i}+H^{AJC}_{i}$. Indeed,
for particular values of the involved parameters
we obtain table 1:\\
\begin{center}
\begin{tabular}{|c|c|c|c|c|}
   \hline
   $i$ &  $\delta_i$ & $\phi$ & $\varphi$ & $H_{i}^{JC}+H_{i}^{AJC}$  \\
   \hline 
   $x $ & $\delta$ & $\frac{3\pi}{2}$ &
$\frac{\pi}{2}$ & $\sqrt{2}\eta\Omega l_{B}\sigma_{x}p_{x}+\hbar\delta\sigma_{z}$  \\
   \hline
   $y $ & $0$ & $\frac{3\pi}{2}$ & $\frac{3\pi}{2}$ & $\sqrt{2}\eta\Omega l_{B}\sigma_{y}p_{y}
   +\frac{\sqrt{2}\hbar\eta\Omega}{l_{B}} 
   \frac{c E}{B}\sigma_{y}y$ \\
   \hline
   $y $ & $0$ & $\frac{3\pi}{2}$ &
$\frac{\pi}{2}$ & $-\frac{\sqrt{2}\hbar\eta\Omega}{l_{B}}\sigma_{x}y$ \\
  \hline
\end{tabular}\\
\vspace{0.5cm}
{\sf Table 1: Linear combinations of JC and AJC Hamiltonian's for given values
of the involved parameters.}
\end{center}
Now by summing all terms in table 1, one gets
\begin{eqnarray}\label{29}
\sum_i\left(H^{JC}_{i}+H^{AJC}_{i}\right)=\sqrt{2}\eta\Omega
l_{B}\left(\sigma_{x}p_{x}+\sigma_{y}p_{y}\right)
+\frac{\sqrt{2}\hbar\eta\Omega}{l_{B}}\left(-\sigma_{x}+ 
\frac{c E}{B}\sigma_{y}\right)y+
\hbar\delta\sigma_{z}.
\end{eqnarray}
A comparison to the Dirac  Hamiltonian (\ref{3h}) yields
to the following correspondence
\begin{eqnarray}\label{o}
\left\{%
\begin{array}{ll}
   \omega_{c}:=2\eta\Omega \\
   \frac{1}{2}g\mu_BB:=\hbar\delta\\
\end{array}%
\right.
\end{eqnarray}
where $\Omega=\Omega_{i}$ and $\eta=\eta_{i}$, $\forall i=x,y$.
(\ref{o}) means that if the Lamb-Dicke parameter $\eta$
is restricted to the ratio between the cyclotron frequency $\om_c$ and
the excitation coupling strength $\Omega$ supplemented by
a variation of magnetic field like  the excitation coupling strength $\del$,
then the present system can be linked with the atoms in quantum optics.
Otherwise, (\ref{o}) tells us that
it is possible to reproduce the 2D Dirac Hamiltonian (\ref{3h}),
with all its quantum relativistic effects, in a controllable
system as a single trapped ion. This conclusion is deduced also by analyzing
massive Dirac fermions in magnetic field in the same framework~\cite{11}.

We close this section by showing that the system under consideration can be reduced to
spintronic ones. In doing so, let us
choose different parameters as 
summarized in table 2:
\\
\begin{center}
\begin{tabular}{|c|c|c|c|c|}
   \hline
   $i$ & $\delta$ & $\phi$ & $\varphi$ &  $H_{i}^{JC}+H_{i}^{AJC}$  \\
   \hline 
     $x$ & $\delta$ & $0$ & $\pi$ & $\sqrt{2}\eta\Omega l_{B}\sigma_{y}p_{x}
   +\hbar\delta\sigma_{z}$ \\
  \hline
   $y$ & $0$ & $0$ & $0$ & $-\sqrt{2}\eta\Omega l_{B}\sigma_{x}p_{y}-\frac{\sqrt{2}\hbar\eta\Omega}{l_{B}}
   \frac{c E}{B}\sigma_{x}y$  \\
  \hline
\end{tabular}
\\
\vspace{0.5cm}
{\sf Table 2: Configuration of four parameters leading to new form
of  $H_{i}^{JC}+H_{i}^{AJC}$.}
\end{center}
Summing  over $i$ to obtain the Hamiltonian
\begin{eqnarray}\label{29}
\sum_i\left(H^{JC}_{i}+H^{AJC}_{i}\right)=\sqrt{2}\eta\Omega
l_{B}\left(\sigma_{y}p_{x}-\sigma_{x}p_{y}\right)
-\frac{\sqrt{2}\hbar\eta\Omega}{l_{B}} 
\frac{c E}{B}\sigma_{x}y+\hbar\delta\sigma_{z}.
\end{eqnarray}
 By taking $E=0$ and $\delta=0$ in (\ref{29}), the resulted
term is nothing but 
the Rashba spin-orbit coupling~\cite{12} where the coefficient $\sqrt{2}\eta\Omega
l_{B}$ is identified
to the coupling parameter of Rashba type.

One can ask for other configurations of the involved parameters and
see what we gain. Indeed, let us take the choice listed in table 3:\\
\begin{center}
\begin{tabular}{|c|c|c|c|c|}
   \hline
    $i$ & $\delta$ & $\phi$ & $\varphi$ &  $H_{i}^{JC}+H_{i}^{AJC}$  \\
   \hline 
     $x$ & $\delta$ & $\frac{3\pi}{2}$ & $\frac{\pi}{2}$ & $\sqrt{2}\eta\Omega l_{B}\sigma_{x}p_{x}
   +\hbar\delta\sigma_{z}$ \\
  \hline
   $y$ & $0$ & $\frac{\pi}{2}$ &
$\frac{\pi}{2}$ & $-\sqrt{2}\eta\Omega l_{B}\sigma_{y}p_{y}-\frac{\sqrt{2}\hbar\eta\Omega}{l_{B}}
\frac{c E}{B}\sigma_{y}y$  \\
  \hline
\end{tabular}
\\
\vspace{0.5cm}
{\sf Table 3: Configuration of four parameters leading to new form
of  $H_{i}^{JC}+H_{i}^{AJC}$.}
\end{center}
As before we sum up to obtain the Hamiltonian 
\begin{eqnarray}\label{292}
\sum_i\left(H^{JC}_{i}+H^{AJC}_{i}\right)=\sqrt{2}\eta\Omega
l_{B}\left(\sigma_{x}p_{x}-\sigma_{y}p_{y}\right)
-\frac{\sqrt{2}\hbar\eta\Omega}{l_{B}}
\frac{c E}{B}\sigma_{y}y+\hbar\delta\sigma_{z}
\end{eqnarray}
which reduces to the linear Dresselhaus coupling~\cite{12}
for $E=0$ and $\delta=0$. According to (\ref{29}) and (\ref{292}), one
can conclude that the system under consideration is also sharing some features with spintronic systems.
Therefore one can go further to investigate the basic features of  (\ref{29}) and (\ref{292}) to extract
more information about these relationships.

\section{ Rabi oscillations} 

The Rabi oscillations of fermions in graphene have been studied in different contexts
to deal with some issues~\cite{ziegler,mishchenko}. 
For example in~\cite{ziegler}
the relation between the canonical model of quantum optics, JC model
and Dirac fermions in quantizing magnetic field was investigated.
It was demonstrated that Rabi oscillations are
observable in the optical response of graphene, providing  with a transparent picture about the structure
of optical transitions. While the longitudinal conductivity reveals chaotic Rabi oscillations, the Hall
component measures coherent ones. This opens up the possibility of investigating a microscopic model of
a few quantum objects in a macroscopic experiment with tunable parameters.

Motivated by the above developments, we analyze the Rabi oscillations of the system under
consideration. We start by
expanding  a general state on to the obtained eigenspinors (\ref{16}) to 
study the corresponding dynamics.
Let us assume that initially
at $t=0$ the atom is in the lower state $\mid f\rangle$ and the
cavity contains precisely $n-1$ photons, i.e. the field is in a
number (Fock) state $\mid n-1\rangle$ with $n = 1, 2, \cdots$.
Then the initial state of the
system is $\mid f, n-1\rangle$, while the interaction term of the
Hamiltonian (\ref{33}) connects this initial state to the final
state $\mid e, n\rangle$. To proceed further, we consider the initial state for the field as
$\mid \psi(0)\rangle=\mid f, n-1, \downarrow\rangle$ and assume an
atom in the excited state is injected into the field.
Using  (\ref{aaa}), (\ref{bbb}) and (\ref{27}), one can write the initial
state as
\begin{eqnarray}\label{ccc}
\mid \psi(0)\rangle= iu_{n}^{-}\mid\psi_{n,\sigma,\va_{n}^+}\rangle-iu_{n}^{+}\mid\psi_{n,\sigma,\va_{n}^-}\rangle.
\end{eqnarray}
Its evolution state can easily be obtained by acting as
\begin{eqnarray}\label{24}
\mid \psi(t)\rangle=e^{-\frac{i}{\hbar}H_{D}t}\ \mid \psi(0)\rangle
=C_{e,n}(t)\mid e,n,\uparrow\rangle+C_{f,n-1}(t)\mid f,n-1,\downarrow\rangle
\end{eqnarray}
where the coefficients 
are given by
\begin{eqnarray}
C_{e,n}(t)=\sqrt{\frac{4\zeta n}{1+4\zeta n}}\sin(\omega_{n}t) \ \ \ \ \ \ \ \ \ \ \ \ \ \ \\
C_{f,n-1}(t)=\cos(\omega_{n}t)+\frac{i}{\sqrt{1+4\zeta
n}}\sin(\omega_{n}t).
\end{eqnarray}
It is easy to check that they
verify the normalization condition
$|C_{e,n}(t)|^{2}+|C_{f,n-1}(t)|^{2}=1$. The frequency of oscillations is given by
\beq
 \omega_{n}:=\frac{E_{n}}{\hbar}=\delta\sqrt{1+4\zeta n}
\eeq
where the parameter
$\zeta=\left(\frac{\omega_{c}}{2\delta}\right)^{2}$ controls the
nonrelativistic limit. (\ref{24}) shows very clearly the
non-stationary states of localization ($\mid e,n,\uparrow\rangle$, $\mid f,n-1,\downarrow\rangle$), which is
nothing but
the Rabi oscillation that is a simple quantum beat interstate $\mid e,n,\uparrow\rangle$
and $\mid f,n-1,\downarrow\rangle$. Moreover, the dynamics of (\ref{24})  is completely similar to
the atomic Rabi oscillations that can be seen from the JC model. This similarity shows another way
how the system under consideration can be linked with Rabi oscillations.

To understand better the present situation, let us analyze the probabilities
of existence. Indeed, for a atom in the state $\mid f,n-1,\downarrow\rangle$, we find
\begin{eqnarray}\label{38}
P_{n-1}^{f}(t)=1-\frac{4\xi n}{1+4\xi n}\sin^{2}(\omega_{n}t)
=1-\left(1-\frac{\delta^{2}}{\omega_{n}^{2}}\right)\sin^{2}(\omega_{n}t)
\end{eqnarray}
where $\om_n$ is now identified to the Rabi frequency $\omega_{R}$. It is clear that,
$P_{n-1}^{f}(t)$
oscillates between a
maximum of unity and a minimum of
$\left(\frac{\delta}{\omega_{n}}\right)^{2}$. (\ref{38}) predicts
sinusoidal Rabi oscillations much like the classical case.
However, even when the field is initially in vacuum state, i.e.
$n= 0$, the probability exists.

Now let see what happens when the atom is in the state $\mid e,n,\downarrow\rangle$. Indeed, after some
algebra
we get the probability 
\begin{eqnarray}
P_{n}^{e}(t)=\frac{4\zeta n}{1+4\zeta
n}\sin^{2}(\omega_{R}t)=\left(1-\frac{\delta^{2}}{\omega_{R}^{2}}\right)\sin^{2}(\omega_{R}t).
\end{eqnarray}
This can be plotted as shown in {Figure 1} for some particular value of $\del$.
One can easily notices that  for the resonance case $ (\delta=0)$, $P_{n}^{e}(t)$  reduces to
\begin{eqnarray}
P_{n}^{e}(t)=\sin^{2}(\omega_{R}t)
\end{eqnarray}
which is in agreement with blue figure (Figure 1). For
$t=\pi/2\omega_{R}$ all the atomic population has
been transferred to the excited state. Clearly, the probabilities
$P_{n}^{e}(t)$ and $P_{n-1}^{f}(t)$ oscillate in time with the
Rabi frequency $\omega_{R}$ where their amplitude 
are maximum at $\delta=0$
 and decrease rapidly as $\delta$ increases. Contrary,
in a stationary state the probabilities of localization of the
particle are equal and constant.

\begin{center}
  \includegraphics[width=4in]{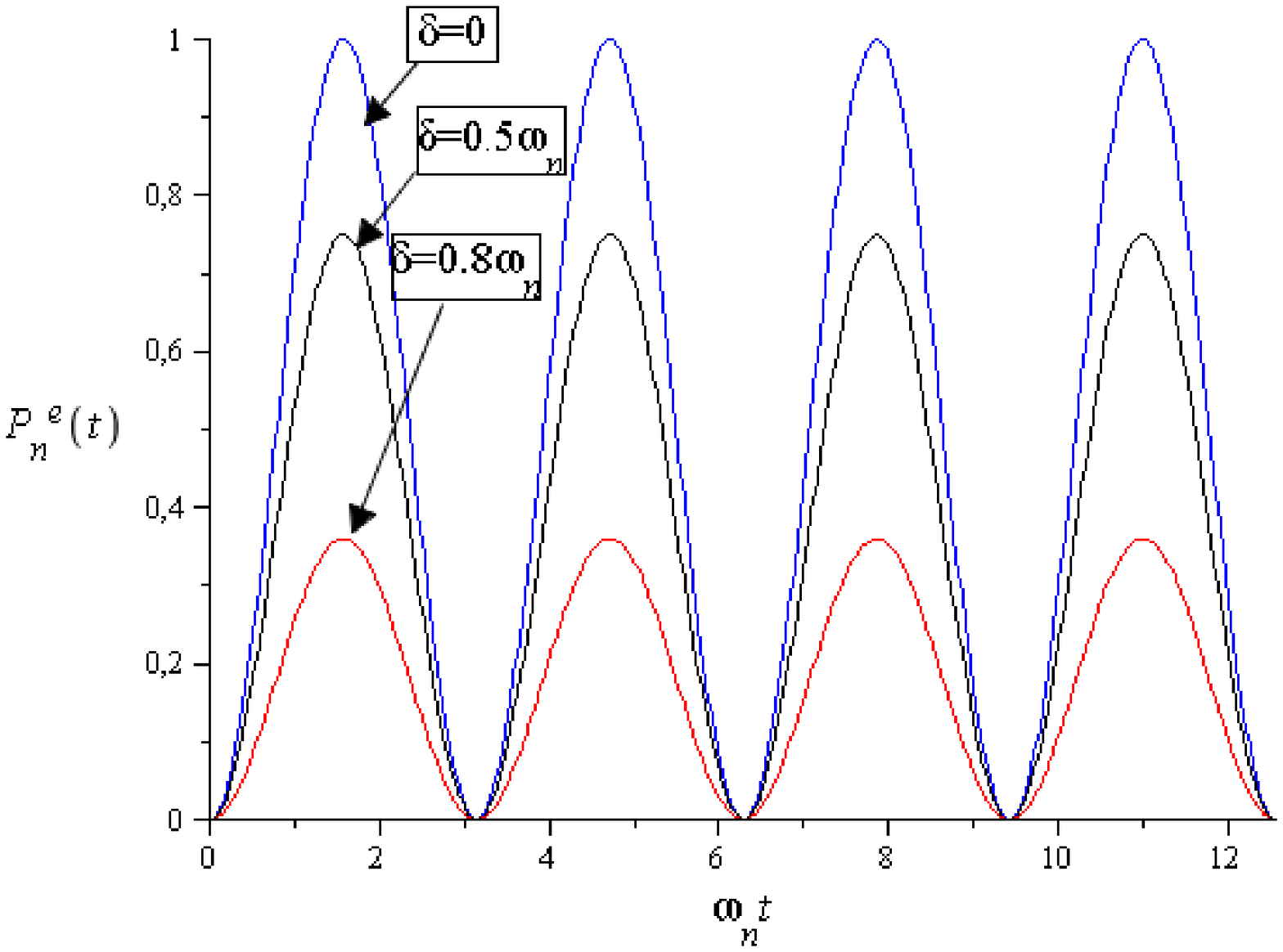}\\
{Figure 1}:  {Plots of $P_{n}^{e}(t)$ versus $t$ for various detunings
$\delta$.}
\end{center}

It is frequently convenient to consider 
the atomic inversion $W(t)$, which is defined as the difference in the excited
and ground state populations. This is
\begin{eqnarray}
W(t)=P_{n}^{e}(t)-P_{n-1}^{f}(t)= 2\left(1-\frac{\delta^{2}}{\omega_{R}^{2}}\right)\sin^{2}(\omega_{R}t) -1
\end{eqnarray}
which can be simplified by assuming that
the atom is initially in the ground state and requiring the resonant case. These allow us to
obtain
\begin{eqnarray}
W(t)=\sin^{2}(\omega_{R}t)-\cos^{2}(\omega_{R}t)=-\cos(2\omega_{R}t)
\end{eqnarray}
where the Rabi frequency reduces now to the quantity
$\omega_{R}=\sqrt{n}\omega_{c}$. Again, for
$t=\pi/2\sqrt{n}\omega_{c}$ all the population is transferred to
the excited state and we have $W(\pi/2\sqrt{n}\omega_{c})=1$. On the other
hand, if we choose $t=\pi/4\sqrt{n}\omega_{c}$, then
$W(\pi/4\sqrt{n}\omega_{c})=0$ and the population is shared
coherently between the excited and ground states. This case shows the constraint 
\beq
C_{e,n}(\pi/4\sqrt{n}\omega_{c})=
C_{f,n-1}(\pi/4\sqrt{n}\omega_{c})=\frac{1}{\sqrt{2}}
\eeq
and therefore the corresponding state is
\begin{eqnarray}\label{sc}
|\psi(\pi/4\sqrt{n}\omega_{c})\rangle=\frac{1}{\sqrt{2}}\left(|e,n,\uparrow\rangle+|f,n-1,\downarrow\rangle\right).
\end{eqnarray}
In summary, one  can see that the behavior of  two-level atom 
interacting with a single electromagnetic mode is surprisingly
rich. In fact, there are  Rabi oscillations that collapse and remain
quiescent, revive, then collapse again.

\section{Zitterbewegung effect}

The Zitterbewegung (ZB) effect, called also trembling motion, is 
consisting of a helicoidal motion of a free Dirac particle. It is a
natural consequence of the non-commutativity of its velocity
operator components $v_F\vec{\sigma}_{j}$, with $i = x, y$.
{In connection with quantum optics realm, this quantum relativistic effect was
studied by Lamata {\it et al.} \cite{lamata1}. The ZB effect}
 has been also studied for different
reasons
in systems made of graphene~\cite{katsnelson,rusin,goerbig}. For instance in \cite{rusin},
the electric current and spatial displacement due the ZB  effect of electrons in
graphene in the presence of an external magnetic field were described. 
It was shown that, when the electrons are prepared in the
form of wave packets, the presence of a quantizing magnetic field $B$ has very important effects on ZB. (1) For
$B\neq 0$ the ZB oscillations are permanent while for $B=0$ they are transient. (2) For $B\neq 0$ many ZB frequencies
appear while for $B=0$ only one frequency is at work. (3) For $B\neq 0$ both interband and intraband (cyclotron)
frequencies contribute to ZB while for $B=0$ there are no intraband frequencies. (4) Magnetic field intensity
changes not only the ZB frequencies but the entire character of ZB spectrum. An emission of electromagnetic
dipole radiation by the trembling electrons was proposed and described. It was argued that graphene in a magnetic
field is a promising system for an experimental observation of Zitterbewegung.

Let us study the ZB effect for massless Dirac fermions in an electromagnetic field
and discuss what will add an electric field to the results presented in~\cite{rusin}.
In doing so, we describe the dynamics of the system under consideration
through the Heisenberg picture. Indeed,
we  calculate the time evolution of the Dirac position operator
\beq
\vec{x}(t)=e^{-\frac{i}{\hbar}H_{D}t}\ \vec{x}\ e^{\frac{i}{\hbar}H_{D}t}
\eeq
which satisfies the Heisenberg equation of motion
\begin{eqnarray}
\frac{d\vec{x}}{dt} =\frac{i
}{\hbar}[H_{D},\vec{x}]=v_F\vec{\sigma}_{j}
\end{eqnarray}
where $H_D$ is the Hamiltonian given in  (\ref{1}). To get the explicit form of
$\vec{x}(t)$, we introduce the dynamics of the Pauli operators
\beq
-i\hbar \frac{d\vec{\sigma}_{j}}{dt}=[H_{D},\vec{\sigma}_{j}]
\eeq
which can be evaluated by  using
the anti-commutation relations
$\{\sigma_{i},\sigma_{j}\}=2\delta_{ij}\mathbb{\mathbb{I}}$
$(i, j=x, y, z)$. This leads to the equation
\begin{eqnarray}\label{g}
-i\hbar \frac{d\vec{\sigma}_{j}}{dt} 
=-2v_F\vec{\pi}_{j}-2eEy\delta_{yj}+2H_{D}\vec{\sigma}_{j}.
\end{eqnarray}
It is convenient for later use to introduce the compact form
\begin{eqnarray}
-i\hbar \frac{d\vec{\sigma}_{j}}{dt}=2H_{D}\vec{\eta}
\end{eqnarray}
where the operator $\eta$ is given by
\begin{eqnarray}\label{h}
\vec{\eta}=\vec{\sigma}_{j}-v_FH_{D}^{-1}\vec{\pi}_{j}-eEyH_{D}^{-1}\delta_{yj}
\end{eqnarray}
and satisfies {the following dynamics}
\begin{eqnarray}
-i\hbar \frac{d\vec{\eta}_{j}}{dt}=-i\hbar
\frac{d\vec{\sigma}_{j}}{dt}=2H_{D}\vec{\eta}.
\end{eqnarray}
It can be solved easily to find
\begin{eqnarray}\label{333}
\vec{\eta}(t)=e^{\frac{2i}{\hbar}H_{D}t}\vec{\eta}_{0}
\end{eqnarray}
where
$\vec{\eta}_{0}$ is a constant operator
\beq
\vec{\eta}_{0}=\vec{\eta}(0)=\vec{\sigma}_{j}(0)-v_FH_{D}^{-1}\vec{\pi}_{j}-eEyH_{D}^{-1}\delta_{yj}.
\eeq
Since we have 
$\{H_{D},\vec{\eta}\}=0=\{H_{D},\vec{\eta}_{0}\}$, thus
(\ref{333}) rewrites as
\begin{eqnarray}\label{j}
\vec{\eta}(t)=\vec{\eta}_{0}e^{\frac{2i}{\hbar}H_{D}t}.
\end{eqnarray}
Combining (\ref{g}), (\ref{h}) and (\ref{j}), to end up with
\begin{eqnarray}\label{}
\frac{d\vec{x}}{dt} 
=v_F^{2}H_{D}^{-1}\vec{\pi}_{j}+eEv_FyH_{D}^{-1}\delta_{yj}+v_F\vec{\eta}_{0}e^{-\frac{2i}{\hbar}H_{D}t}
\end{eqnarray}
which can be integrated  to get  the time evolution
\begin{eqnarray}\label{x}
\vec{x}(t)=v_F^{2}H_{D}^{-1}\vec{\pi}_{j}t+eEv_FyH_{D}^{-1}\delta_{yj}t+\frac{1}{2}i\hbar
v_F\vec{\eta}_{0}H_{D}^{-1}e^{-\frac{2i}{\hbar}H_{D}t}+\vec{x}(0)
\end{eqnarray}
where $\vec{x}(0)$ is a constant (operator) of integration
\begin{eqnarray}\label{}
\vec{x}(0)-\frac{1}{2}i\hbar
v_F\vec{\sigma}_{j}(0)H_{D}^{-1}+\frac{1}{2}i\hbar
v_F^{2}H_{D}^{-2}\vec{\pi}_{j}+\frac{1}{2}i\hbar
v_FeEyH_{D}^{-2}\delta_{yj}.
\end{eqnarray}
Now using the correspondence (\ref{o}), which is linking our system with  atoms in quantum optics,
we map (\ref{x}) as
\begin{eqnarray}\label{59}
\vec{x}(t) &=&\vec{x}(0)+\frac{2\eta^{2}l_{B}^{2}\Omega^{2}\vec{\pi}_{j}+\sqrt{2}eE\eta l_{B}\Omega y\delta_{yj}}{H_{D}}t \\
&+&\left(\vec{\sigma}_{j}(t)-\frac{\sqrt{2}\eta
l_{B}\Omega\vec{\pi}_{j}+eEy\delta_{yj}}{H_{D}}\right)\frac{i\sqrt{2}\hbar\eta
l_{B}\Omega}{2H_{D}}\left(1-e^{\frac{2i}{\hbar}H_{D}t}\right).\nonumber
\end{eqnarray}
This result is showing different contributions originated from {different sources}. Indeed,
the first two terms on the r.h.s. account for the classical
kinematics of a free particle in electromagnetic field $(\vec E, \vec B)$. However, the last term represents a
quantum oscillation about the classical motion with a frequency
$\frac{2|E_{D}|}{\hbar}$ where $E_{D}=\va_n $. This oscillating term is the so-called ZB effect 
or "trembling
motion" and its
frequency associated with the measurable quantity $\langle
\vec{x}(t)\rangle$ can be estimated as
\begin{eqnarray}\label{}
\omega_{ZB}=\frac{2|E_{D}|}{\hbar}=2\sqrt{4\eta^{2}\Omega^{2}|n|+\delta^{2}}
\end{eqnarray}
where the correspondence (\ref{o}) is used.
On the other hand, the presence of the electric field $\vec E$   is clearly shown in equation (\ref{59}). Thus,
one can  interpret  the terms $\left(\sqrt{2}eE\eta l_{B}\Omega y\delta_{yj}\right)$ and $\left(eEy\delta_{yj}\right)$ as shift
with respect to {the original terms, which of course disappear  if we require} $\vec E=0$.
 This tells us that
{our results are interesting and generals such that one can recover
the standard ones by simply fixing some parameters.}


\section{Conclusion}

The solutions of the energy spectrum of a system
of massless Dirac fermions in the electromagnetic field
are obtained. The corresponding Dirac Hamiltonian $H_D$ is
used to investigate two-level atom
interacting with a single electromagnetic field.
The connection between JC model in quantum optics with $H_D$
is established  in
a simple manner without taking any limit on the strength of the
electromagnetic field. This can be served as a tool  to study atomic transitions
in quantum optics using the relativistic quantum mechanical models
in the presence of the electromagnetic field.

More precisely, after mapping the  eigenspinors in terms of ground and excite
states of two-level atom,
we  derived a new form of the Dirac Hamiltonian. This is used to
establish a link 
with JC and AJC models as linear combination under the correspondence between their
parameters. Making a suitable choose of the involved parameters in the game,
we showed that the linear combination can be reduced either to
the Rashba or Dresselhous
couplings. This is a significant results because it relies two sectors such
graphene and quantum optic systems.

Because of the Rabi oscillations are occurring in the  JC model and its relation to
our system, we analyzed them.
In fact,
we mapped the eigenspinors in terms of
two level atom states, i.e. ground and excited and showed that
the Rabi frequency can be obtained in terms of the quantized energy levels.
Calculating the probabilities for atom in ground and excited states, separately,  we obtained different
results showing that these probabilities are maximum when the detuning $\delta$ is null and
decrease rapidly as $\delta$ increases. These are used to discuss the resonant case
by evaluating the atomic inversion $W(t)$ and  conclude that the population of the state is
depending on the value attributed to $W(t)$.

Finally, we analyzed the Zeitterbewegung effect for massless Dirac fermions in
the electromagnetic field. 
This was done by using the Heisenberg picture
dynamics to evaluate the time evolution of  the Dirac position operator. This allowed us to
obtain a quantum oscillation about the classical motion with a defined
frequency. Its
oscillation between the negative and positive energy states is
similar to the Rabi oscillation in the two-level atom. By
evaluating the position average, we obtained the ZB frequency in terms of the quantized
energy system.

{We close this part by listing some ideas, which can be worked by adopting our approach
presented in this paper.
Indeed, it will be of interest to analyze
the massless
Dirac fermions in the presence of a perpendicular magnetic field
and by considering an electric field supported by $2\times 2$ unit
matrix as formulated for graphene in \cite{goerbig1}. This certainly will
bring new results and interesting conclusions due to the reduction of the magnetic field from
$B$ to $B'=B\sqrt{1-\left(\frac{E}{V_FB}\right)^2}$. One also can study in
this framework the  tilted magnetic field case
instead of an uniform one}.

\section*{Acknowledgment}

The generous support provided by the Saudi Center for Theoretical Physics (SCTP) is
highly appreciated by AJ and AEM.  AJ
acknowledges partial support by King Faisal University.
{The authors are indebted to the referees for their instructive comments.}


\end{document}